\documentstyle[12pt]{article}

\newlength{\minitwocolumn}
\catcode`\@=11
\def\relaxnext@{\let\next\relax}
\font\tenmsy=msym10 scaled\magstep1
\font\sevenmsy=msym7 scaled\magstep1
\font\fivemsy=msym5  scaled\magstep1
\newfam\msyfam
\textfont\msyfam=\tenmsy
\scriptfont\msyfam=\sevenmsy
\scriptscriptfont\msyfam=\fivemsy
\font\teneuf=eufm10 scaled\magstep1
\font\seveneuf=eufm7 scaled\magstep1
\font\fiveeuf=eufm5 scaled\magstep1
\newfam\euffam
\textfont\euffam=\teneuf
\scriptfont\euffam=\seveneuf
\scriptscriptfont\euffam=\fiveeuf
\def\frak{\relaxnext@\ifmmode\let\next\frak@\else
 \def\next{\Err@{Use \string\frak\space only in math mode}}\fi\next}
\def\goth{\relaxnext@\ifmmode\let\next\frak@\else
 \def\next{\Err@{Use \string\goth\space only in math mode}}\fi\next}
\def\frak@#1{{\frak@@{#1}}}
\def\frak@@#1{\noaccents@\fam\euffam#1}
\def\Bbb{\relaxnext@\ifmmode\let\next\Bbb@\else
 \def\next{\Err@{Use \string\Bbb\space only in math mode}}\fi\next}
\def\Bbb@#1{{\Bbb@@{#1}}}
\def\Bbb@@#1{\noaccents@\fam\msyfam#1}
\def\accentfam@{7}
\def\noaccents@{\def\accentfam@{0}}
\catcode`\@=\active
\newcommand{\bz}{{\Bbb Z}}

\newcommand{\bc}{{\Bbb C}}
\makeatletter
\@addtoreset{equation}{section}
\makeatother

\begin{document}
\begin{flushright}
RIMS-930 \\
Aug. 1993
\end{flushright}
\vspace{24pt}
\begin{center}
\begin{Large}
{\bf Spontaneous Polarization of the
$\Bbb Z_{n}$-Baxter Model}
\end{Large}

\vspace{36pt}

Yas-Hiro Quano\raisebox{2mm}{$\star$}

\vspace{6pt}

{\it Research Institute for Mathematical Sciences}\\
{\it Kyoto University, Kyoto 606, Japan}

\vspace{72pt}

\underline{Abstract}
\end{center}

\vspace{6pt}
We show that correlation functions of
the $\bz _n $-Baxter model in the principal regime
satisfy a system of difference equations.
We obtain the spontaneous polarization
of the $\bz _n $-Baxter model
as a solution of the simplest difference equation.

\vspace{24pt}

\vfill
\hrule

\vskip 3mm
\begin{footnotesize}

\noindent\raisebox{2mm}{$\star$}A Fellow of
the Japan Society for the Promotion of Science for
Japanese Junior Scientists.
Partly supported by the Grant-in-Aid for Scientific Research
{}from the Ministry of Education, Science and Culture (No. 04-2297).
\end{footnotesize}
\newpage

\section{Introduction}

It is widely recognized that
systems of holonomic difference equations
commonly appear in integrable quantum field theories
and solvable lattice statistical models.
Smirnov \cite{FF} found that form factors in
two dimensional integrable massive models
satisfy difference equations.
Frenkel and Reshetikhin \cite{FR}
showed that
correlation functions of the intertwining operators
of the quantum affine algebra obey difference equations,
and that the resulting connection matrices give
elliptic solutions \cite{ABF,JMO} to the Yang-Baxter equation (YBE).
In \cite{DFJMN,JMMN} it was proposed that
correlations of the $XXZ$ model in the anti-ferromagnetic regime
can be formulated in terms of the trace of products of
the intertwining operators
of $U_{q}({\hat{\frak s \frak l _2 }})$.
The higher spin analog of the $XXZ$ model \cite{IIJMNT} and
the higher rank generalization \cite{DO,Ko} were also studied.
The integral formula of correlation functions
of the $XXZ$ model is given
in \cite{JMMN}.
That of
the matrix element
of product of intertwining operators
of $U_{q}({\hat{\frak s \frak l _2 }})$
for arbitrary level is obtained
in \cite{KQS}.

It was remarked in \cite{FR} that
even if the trigonometric $R$-matrices is replaced by
the elliptic ones, integrability of difference equations survives.
Jimbo, Miwa and Nakayashiki \cite{JMN} actually obtained
difference equations for correlations of the eight-vertex model
in the principal regime.
In the present paper we generalize their results
to the case of the $\bz _n $-Baxter model \cite{Zn}.

The $\bz _n $-Baxter model is
a vertex model on a two-dimensional square lattice ${\cal L}$
such that the state variables take on $\bz _n $-spin.
Each oriented line of ${\cal L}$
carries a spectral parameter varying from line to line.
We assign $\bz _n $-valued local states on edges.
Set

\begin{minipage}{\minitwocolumn}
\unitlength 1mm
\begin{picture}(40,30)
\put(47,15){$R(z_1 -z_2 )^{ik}_{jl}:=
$}
\end{picture}
\end{minipage}%
\hspace{\columnsep}%
\begin{minipage}{\minitwocolumn}
\unitlength 1mm
\begin{picture}(40,30)
\put(-5,6){\begin{picture}(101,0)
\put(0,10){\vector(1,0){15}}
\put(15,10){\line(1,0){5}}
\put(10,0){\vector(0,1){15}}
\put(10,15){\line(0,1){5}}
\put(-3,9.5){$j$}
\put(9,21){$k$}
\put(9,-4){$l$}
\put(21,9.5){$i$}
\put(13,6){$z_{1}$}
\put(11,14){$z_{2}$}
\end{picture}
}
\end{picture}
\end{minipage}
be a local Boltzmann weight
for a single vertex with bond states
$i, j, k, l \in \bz _n$.
Arrows denotes orientations of lines.
In this case the weights are given as follows \cite{RT}:
\begin{equation}
R(z)^{ik}_{jl}=\left\{ \begin{array}{ll}
\displaystyle\frac{h(z)\theta ^{(i-k)}(z+w)}
{\theta ^{(j-k)}(z)\theta ^{(i-j)}(w)} &
\mbox{if $i+k=j+l$, mod $n$,} \\
0 & \mbox{otherwise,} \end{array} \right. \label{Bel}
\end{equation}
where
$$
\theta^{(j)}(z):=
\displaystyle\sum_{m\in \bz }
\exp \left\{ \pi \sqrt{-1}
(m+\displaystyle\frac{1}{2}-\displaystyle\frac{j}{n})^{2} n\tau
+2\pi \sqrt{-1}
(m+\displaystyle\frac{1}{2}-\displaystyle\frac{j}{n})
(z+\displaystyle\frac{1}{2}) \right\},
%\label{Rieth}
$$
$$
h(z):=\prod_{j=0}^{n-1} \theta^{(j)}(z)/
\prod_{j=1}^{n-1} \theta^{(j)}(0),
$$
and $w\neq 0~\mbox{mod~}\bz +\bz \tau $, is a constant.
Now we assume that $0<t<q<u<1$, where
$t:=\exp (\pi\sqrt{-1}\tau ),
 q:=\exp (\pi\sqrt{-1}w)$, and $u:=\exp (-\pi\sqrt{-1}z)$.
Then the elliptic theta functions are expressed
in terms of the product series
\begin{equation}
\begin{array}{c}
\theta^{(j)}(z)=
\sqrt{-1}\omega ^{j/2}t^{n(1/2-j/n)^{2}}
u^{-1+2j/n}
(t^{2n}; t^{2n})_{\infty}
(t^{2j}u^{2}; t^{2n})_{\infty}
(t^{2(n-j)}u^{-2}; t^{2n})_{\infty}, \\
h(z)=\sqrt{-1}t^{n/4} \displaystyle\frac{(t^{2n}; t^{2n})^{3}_{\infty}}
                                        {(t^{2}; t^{2})^{2}_{\infty}}
u^{-1}
(u^{2}; t^{2})_{\infty}
(t^{2}u^{-2}; t^{2})_{\infty},
\end{array}\label{fpro}
\end{equation}
where
$$
(a; q_{1}, \cdots , q_{k} )_{\infty } :=
\prod_{m_{1}=0}^{\infty} \cdots
\prod_{m_{k}=0}^{\infty} (1-aq_{1}^{m_{1}}\cdots q_{k}^{m_{k}}).
$$
Following Baxter \cite{ESM}
we call such domain of parameters the principal regime.
Note that (\ref{Bel}) is weights of the eight-vertex model when $n=2$.

The plan of this paper is as follows.
In section 2
we summarize our result on
the unitarity and the crossing symmetry of
the $\bz _n $-Baxter model \cite{H,QD}.
In section 3 we introduce the normalized
$R$-matrices by the partition function
per site in the thermodynamic limit.
In section 4 we show that correlation functions defined
by using this normalized $R$-matrices
satisfy $q$-difference equations.
In section 5 solving the simplest one
we obtain the spontaneous polarization.
By setting $n=2$, our result
reproduces that of \cite{JMN}.
We can also get Koyama's result \cite{Ko} by taking the limit
$t\rightarrow 0$.
In section 6 we give some remarks.

\section{Unitarity and Crossing Symmetry}

Let $V=\bc ^n $ and $\{v_i \}_{i\in \bz _n }$ be
the standard orthonormal basis of $V$.
Then the $R$-matrix $R(z)$
whose $(ik,jl)$-element is
$R(z)^{ik}_{jl}$ gives a linear map
$V\otimes V \rightarrow V\otimes V$
\begin{eqnarray*}
R(z)(v_j \otimes v_ l )=\sum_{i, k \in \bz _n }
(v_i \otimes v_k ) R(z)^{ik}_{jl}.
\end{eqnarray*}
The $\bz _n $-Baxter model has the $\bz _n $-symmetry
\begin{equation}
\begin{array}{cl}
\mbox{({\romannumeral 1})} & R(z)^{ik}_{jl}=0,
\mbox{~~unless $i+k=j+l$,~~mod $n$}, \\
\mbox{({\romannumeral 2})} & R(z)^{i+p,k+p}_{j+p,l+p}=
R(z)^{ik}_{jl},
\mbox{~~for every $i,j,k,l$ and $p\in \bz _n$}.
\end{array} \label{Znsym}
\end{equation}
In terms of two linear map in $V$
\begin{equation}
gv_i =\omega ^i v_i ,~~~~~~
hv_i =v_{i-1}, \label{gh}
\end{equation}
where, $\omega =\exp (2\pi \sqrt{-1}/n),$
(\ref{Znsym}) can be rephrased as follows:
\begin{equation}
\begin{array}{rcl}
R(z)(g\otimes g) & = & (g\otimes g)R(z), \\
R(z)(h\otimes h) & = & (h\otimes h)R(z).
\end{array} \label{gln}
\end{equation}
Owing to (\ref{gln})
the $\bz _n $-Baxter model is called the
$\frak g \frak l _n $-invariant model.
Furthermore, the $R$-matrix satisfies the Yang-Baxter equation
\begin{eqnarray}
R_{1;2}(z_1 -z_2 )
R_{1;3}(z_1 -z_3 )
R_{2;3}(z_2 -z_3 )=
R_{2;3}(z_2 -z_3 )
R_{1;3}(z_1 -z_3 )
R_{1;2}(z_1 -z_2 ), \label{YBE}
\end{eqnarray}
where $R_{i;j}(z)$ denotes the matrix on $V^{\otimes 3}$,
which acts as $R(z)$ on the $i$-th and $j$-th components and
as identity on the other one.
In what follows subscripts of any linear operator
signify the components on which it acts.

Let us reformulate this lattice model
in terms of the scattering theory of elementary particles.
Consider the system consisting of $n$ kinds of particles
with equal mass
and let $i^* $ stand for the antiparticle of the particle $i$.
In a two dimensional integrable field theory,
both rapidities and $\bz _n $-current preserve
before and after any two-body collision.
Let $V=\bc ^n $ and $V_z $ be a copy of $V$ with
a rapidity (spectral) parameter $z$.
Then $R(z_1 -z_2 )$
can be regarded as a linear map
$R^{V_{z_1 }, V_{z_2 }}$
acting on $V_{z_1 }\otimes V_{z_2 }$.
It is due to the Lorentz invariance
that $R^{V_{z_1 }, V_{z_2 }}$ depends only upon
the difference of rapidities $z_1 -z_2 $.
We also set $\check{R}^{V_{z_1 }, V_{z_2 }}=PR^{V_{z_1 }, V_{z_2 }}$,
where $P$ is the permutation
$V_{z_1 }\otimes V_{z_2 } \rightarrow V_{z_2 }\otimes V_{z_1 }$.
In this notation, YBE (\ref{YBE}) reads as follows:
\begin{eqnarray}
(I\otimes \check{R}^{V_{z_1}, V_{z_2}})
(\check{R}^{V_{z_1}, V_{z_3 }}\otimes I )
(I\otimes \check{R}^{V_{z_2}, V_{z_3 }})=
(\check{R}^{V_{z_2}, V_{z_3 }}\otimes I)
(I \otimes \check{R}^{V_{z_1}, V_{z_3 }})
(\check{R}^{V_{z_1}, V_{z_2}}\otimes I), \label{RYBE}
\end{eqnarray}
as a linear map
$V_{z_1 }\otimes V_{z_2 } \otimes V_{z_3 }
\rightarrow
 V_{z_3 } \otimes V_{z_2 }\otimes V_{z_1 }$.

Next we extend this formulation for
arbitrary spaces $K$ and $L$ of tensor product of some $V$'s.
For $K=V_{z_1 }\otimes \cdots \otimes V_{z_k }$ and
    $L=V_{z'_1 }\otimes \cdots \otimes V_{z'_l }$,
it is very natural
to define a linear map $\check{R}^{K,L}:
K \otimes L \rightarrow L \otimes K$
as follows \cite{IVC,H}:
$$
\begin{array}{rl}
\check{R}^{K, V_{z'}}:= &
\check{R}^{V_{z_1 }, V_{z'}}_{1;2}
\cdots
\check{R}^{V_{z_k }, V_{z'}}_{k;k+1}, \\
 \check{R}^{K,L}:= &
\check{R}^{K, V_{z'_l }}_{l\cdots k+l-1; k+l}
\cdots
\check{R}^{K, V_{z'_1 }}_{1\cdots k; k+1}.
\end{array}
$$
YBE holds for $ \check{R}^{K,L}$
by virtue of YBE for $\check{R}^{V,V}$ (\ref{RYBE})
\begin{eqnarray}
(I\otimes \check{R}^{K, L})
(\check{R}^{K, M}\otimes I )
(I\otimes \check{R}^{L, M})=
(\check{R}^{L, M}\otimes I)
(I \otimes \check{R}^{K, M})
(\check{R}^{K, L}\otimes I), \label{KYBE}
\end{eqnarray}
as a linear map
$K\otimes L \otimes M \rightarrow M \otimes L\otimes K$.

Now we summarize the unitarity and the crossing symmetry
of the $\bz _n $-Baxter model.

(1)~~The unitarity
{}~~~~~~The unitarity or the first inversion relation \cite{RT,QD}
is given by
\begin{equation}
\check{R}^{V_{z_1 }, V_{z_2 }}\check{R}^{V_{z_2 }, V_{z_1 }} =
\frac{h(z_1 -z_2 +w)h(-z_1 +z_2 +w)}{h^2 (w)}I\otimes I.
\label{uni}
\end{equation}

(2)~~The crossing symmetry
{}~~~~~~Let $V^* $ be the dual space of $V$ and
$\{v^{*}_i \}_{i\in \bz _n }$ be
the dual basis of $\{v_i \}_{i\in \bz _n }$.
Then we have the isomorphism $C: V^* \rightarrow \Lambda ^{n-1}(V)$
\begin{eqnarray}
Cv^* _i =\displaystyle\sum_{i_1 , \cdots , i_{n-1}}
\frac{\epsilon _i ^{i_1 \cdots i_{n-1}}}{\sqrt{(n-1)!}}
                 v_{i_1 }\otimes
                 \cdots \otimes
                 v_{i_{n-1}},
\label{hom}
\end{eqnarray}
where $\epsilon _i ^{i_1 \cdots i_{n-1}}$ is
the $n$-th order completely antisymmetric tensor.

The $R$-matrices corresponding to the collision between
a particle and an antiparticle is given as follows:
\begin{equation}
\begin{array}{rcl}
\check{R}^{V_{z_1 }, V^{*}_{z_{2}}} & = & (C\otimes I)^{-1}
\check{R}^{V_{z_1 }, V_{z_2 +(n-1)w}\otimes \cdots \otimes V_{z_2 +w}}
(I\otimes C), \label{VV*} \\
\check{R}^{V^{*}_{z_1 }, V_{z_{2}}} & = & (I\otimes C)^{-1}
\check{R}^{V_{z_1 +(n-1)w}\otimes \cdots \otimes V_{z_1 +w}, V_{z_2 }}
(C\otimes I).
\end{array}
\label{pap}
\end{equation}
The matrix elements of these are
\begin{eqnarray*}
\check{R}^{V_{z_1 }, V^{*}_{z_{2}}}(v_j \otimes v^{*}_{l})
& = & \sum_{i^{*},k}(v^{*}_{i}\otimes v_{k})
(\check{R}^{V_{z_1 }, V^{*}_{z_{2}}})^{i^{*}k}_{jl^{*}}, \\
\check{R}^{V^{*}_{z_1 }, V_{z_{2}}}(v^{*}_j \otimes v_{l})
& = & \sum_{i,k^{*}}(v_{i}\otimes v^{*}_{k})
(\check{R}^{V_{z_1 }, V^{*}_{z_{2}}})^{ik^{*}}_{j^{*}l},
\end{eqnarray*}
which meet the crossing symmetry \cite{H,QD}
\begin{equation}
\begin{array}{rcl}
(\check{R}^{V_{z_1 }, V^{*}_{z_{2}}})^{i^* k}_{jl^* } & = &
(\check{R}^{V_{z_2 }, V_{z_1 }})^{kl}_{ij}
\displaystyle\prod_{p=2}^{n-1} \frac{h(-z_1 +z_2 +pw)}{h(w)}, \\
(\check{R}^{V^{*}_{z_1 }, V_{z_{2}}})_{i^* j}^{kl^* } & = &
(\check{R}^{V_{z_2 }, V_{z_1 +nw}})_{jl}^{ik}
\displaystyle\prod_{p=1}^{n-2} \frac{h(-z_1 +z_2 -pw)}{h(w)}.
\end{array}
\label{ncross}
\end{equation}
{}From (\ref{uni})and (\ref{ncross}),
we have the following second inversion relation \cite{RT,QD}
\begin{equation}
\sum_{jl}
\check{R}^{il}_{kj}(z)\check{R}^{k'j}_{i'l}(-z-nw)=
\frac{h(-z)h(z+nw)}{h^2 (w)}
\delta^{i}_{i'} \delta_{k}^{k'}. \label{2nd}
\end{equation}
The $R$-matrix corresponding to
antiparticle-antiparticle scattering is also defined by
\begin{eqnarray*}
\check{R}^{V^{*}_{z_1 }, V^{*}_{z_2 }}=(C\otimes C)^{-1}
\check{R}^{V_{z_1 +(n-1)w}\otimes \cdots \otimes V_{z_1 +w},
           V_{z_2 +(n-1)w}\otimes \cdots \otimes V_{z_2 +w}}
(C\otimes C).
\end{eqnarray*}

\section{Partition Function and $S$-Matrices}

Let $\kappa (z)$ be the partition function
per site in the thermodynamic limit.
With the help of two inversion relations (\ref{uni}), (\ref{2nd})
and Baxter's corner transfer matrix method \cite{ESM}, we can get
the functional equations for $\kappa (z)$:
\begin{equation}
\begin{array}{rcl}
\kappa (z) \kappa (-z) & = &
\displaystyle\frac{h(z+w)h(-z+w)}{h^2 (w)}, \\
\kappa (z) \kappa (-z-nw) & = &
\displaystyle\frac{h(-z)h(z+nw)}{h^2 (w)}.
\end{array}
\label{inv}
\end{equation}
Hereafter $\kappa (z)$ is often denoted by $\kappa (u)$
through the relation $u=\exp (-\pi \sqrt{-1}z)$.

In the principal regime using (\ref{fpro})
the following expression solves (\ref{inv}) \cite{RT}
\begin{equation}
\kappa (u)=u^{-(n-2)/n}
\frac{(u^2 ; t^2 )_{\infty}(t^2 u^{-2} ; t^2 )_{\infty}}
     {(q^2 ; t^2 )_{\infty}(t^2 q^{-2} ; t^2 )_{\infty}}
\bar{\kappa }(u),
\end{equation}
where
$$
\bar{\kappa }(u)=
\frac{(q^2 u^2 ; t^2 , q^{2n})_{\infty}
      (q^{2n} u^{-2} ; t^2 , q^{2n})_{\infty}
      (t^2 q^{-2} u^2 ; t^2 , q^{2n})_{\infty}
      (t^2 q^{2n} u^{-2} ; t^2 , q^{2n})_{\infty}}
     {(q^{2+2n} u^{-2} ; t^2 , q^{2n})_{\infty}
      (u^2 ; t^2 , q^{2n})_{\infty}
      (t^2 q^{-2+2n} u^{-2} ; t^2 , q^{2n})_{\infty}
      (t^2 u^2 ; t^2 , q^{2n})_{\infty}}.
$$
For later convenience, we define $\check{S}^{K,L}=PS^{K,L}$ by
\begin{equation}
\begin{array}{rcl}
\check{S}^{V_{z_1 }, V_{z_2 }} & = &
\kappa (z_1 -z_2 )^{-1}\check{R}^{V_{z_1 }, V_{z_2 }}, \\
\check{S}^{V_{z_1 }, V^{*}_{z_2 }} & = &
\displaystyle\prod_{p=1}^{n-1} \kappa (z_1 -z_2 -pw)^{-1}
\check{R}^{V_{z_1 }, V^{*}_{z_2 }}, \\
\check{S}^{V^{*}_{z_1 }, V_{z_2 }} & = &
\displaystyle\prod_{p=1}^{n-1} \kappa (z_1 -z_2 +pw)^{-1}
\check{R}^{V^{*}_{z_1 }, V_{z_2 }}, \\
\check{S}^{V^{*}_{z_1 }, V^{*}_{z_2 }} & = &
\displaystyle\prod_{p,q=1}^{n-1} \kappa (z_1 -z_2 +(p-q)w)^{-1}
\check{R}^{V^{*}_{z_1 }, V^{*}_{z_2 }}.
\end{array}
\end{equation}
Then the unitarity and the crossing symmetry for $\check{S}^{K,L}$ hold:
\begin{equation}
\begin{array}{rcl}
\check{S}^{K,L}\check{S}^{L,K} & = & \mbox{id.}, \\
(\check{S}^{V_{z_1 }, V^{*}_{z_{2}}})^{i^* k}_{jl^* } & = &
(\check{S}^{V_{z_2 }, V_{z_1 }})^{kl}_{ij}, \\
(\check{S}^{V^{*}_{z_1 }, V_{z_{2}}})_{i^* j}^{kl^* } & = &
(\check{S}^{V_{z_2 }, V_{z_1 +nw}})^{ik}_{jl}.
\end{array} \label{Smat}
\end{equation}
The first one is obvious.
The last two follow from
\begin{eqnarray*}
\kappa (-z)^{-1} \prod_{j=1}^{n-1} \kappa (z-jw) =
\prod_{j=2}^{n-1} \frac{h(-z+jw)}{h(w)},
\end{eqnarray*}
which can be checked by explicit calculation.
The relations (\ref{Smat}) are visualized as

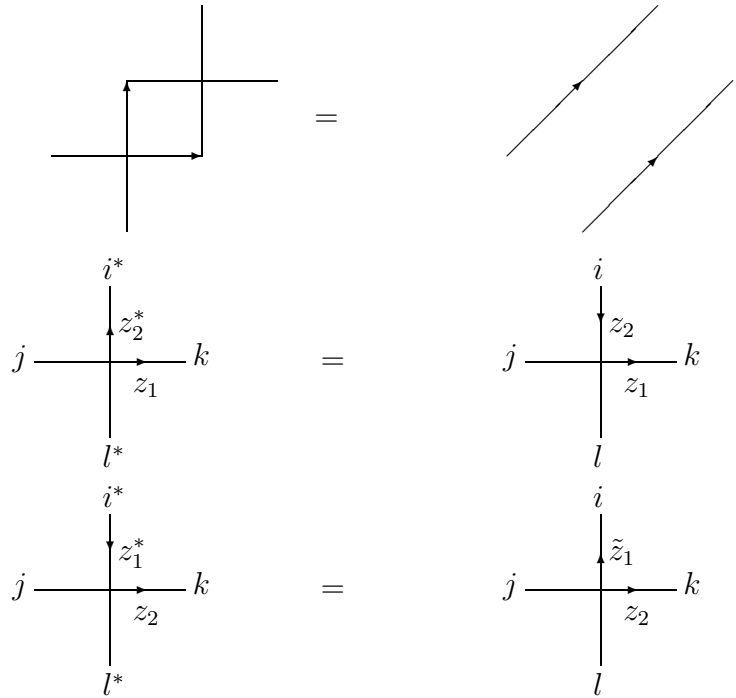
\begin{minipage}{\minitwocolumn}
\unitlength 1mm
\begin{picture}(40,35)
\put(36,1){\begin{picture}(101,0)
\put(0,10){\vector(1,0){20}}
\put(20,10){\line(0,1){20}}
\put(10,0){\vector(0,1){20}}
\put(10,20){\line(1,0){20}}
\put(35,14){$=$}
\end{picture}
}
\end{picture}
\end{minipage}%
\hspace{\columnsep}%
\begin{minipage}{\minitwocolumn}
\unitlength 1mm
\begin{picture}(40,35)
\put(10,1){
\begin{picture}(101,0)
\put(0,10){\vector(1,1){10}}
\put(10,20){\line(1,1){10}}
\put(10,0){\vector(1,1){10}}
\put(20,10){\line(1,1){10}}
\end{picture}
}
\end{picture}
\end{minipage}
\begin{minipage}{\minitwocolumn}
\unitlength 1mm
\begin{picture}(40,30)
\put(40,4){\begin{picture}(101,0)
\put(0,10){\vector(1,0){15}}
\put(15,10){\line(1,0){5}}
\put(10,0){\vector(0,1){15}}
\put(10,15){\line(0,1){5}}
\put(-3,9.5){$j$}
\put(9,21){$i^{*}$}
\put(9,-4){$l^{*}$}
\put(21,9.5){$k$}
\put(13,6){$z_{1}$}
\put(11,14){$z^{*}_{2}$}
\put(38,9){$=$}
\end{picture}
}
\end{picture}
\end{minipage}%
\hspace{\columnsep}%
\begin{minipage}{\minitwocolumn}
\unitlength 1mm
\begin{picture}(40,30)
\put(20,4){\begin{picture}(101,0)
\put(0,10){\vector(1,0){15}}
\put(15,10){\line(1,0){5}}
\put(10,0){\line(0,1){15}}
\put(10,20){\vector(0,-1){5}}
\put(-3,9.5){$j$}
\put(9,21){$i$}
\put(9,-4){$l$}
\put(21,9.5){$k$}
\put(13,6){$z_{1}$}
\put(11,14){$z_{2}$}
\end{picture}
}
\end{picture}
\end{minipage}
\begin{minipage}{\minitwocolumn}
\unitlength 1mm
\begin{picture}(40,30)
\put(40,4){\begin{picture}(101,0)
\put(0,10){\vector(1,0){15}}
\put(15,10){\line(1,0){5}}
\put(10,0){\line(0,1){15}}
\put(10,20){\vector(0,-1){5}}
\put(-3,9.5){$j$}
\put(9,21){$i^{*}$}
\put(9,-4){$l^{*}$}
\put(21,9.5){$k$}
\put(13,6){$z_{2}$}
\put(11,14){$z^{*}_{1}$}
\put(38,9){$=$}
\end{picture}
}
\end{picture}
\end{minipage}%
\hspace{\columnsep}%
\begin{minipage}{\minitwocolumn}
\unitlength 1mm
\begin{picture}(40,30)
\put(20,4){\begin{picture}(101,0)
\put(0,10){\vector(1,0){15}}
\put(15,10){\line(1,0){5}}
\put(10,0){\vector(0,1){15}}
\put(10,15){\line(0,1){5}}
\put(-3,9.5){$j$}
\put(9,21){$i$}
\put(9,-4){$l$}
\put(21,9.5){$k$}
\put(13,6){$z_{2}$}
\put(11,14){$\tilde{z}_{1}$}
\end{picture}
}
\end{picture}
\end{minipage}

\vspace{5mm}
\noindent where $\tilde{z}=z+nw$.
Note that $z$ ($z^{*}$) in the above figure
represents the space $V_{z}$ ($V^{*}_{z}$), while
$i$ ($i^{*}$) denotes $v_{i}$ ($v^{*}_{i}$).

{\bf Remark}~~~~
Speaking in terms of Young tableau,
$V=\Box$, the fundamental representation
of $\frak s \frak l _n $, while
$V^* $ corresponds to $n-1$ vertical $\Box$'s,
the space of antisymmetric tensors in $V^{\otimes n-1}$.
Thus both $V$ and $V^{*}$
are indispensable in our formulation.
For $n=2$ \cite{JMN} from (\ref{hom}) and (\ref{pap}),
one can do without $V^{*}$
by identifying $v^{*}_{i}=(-1)^{i}v_{1-i},
V^{*}_{z}=V_{z+w}$.

\section{Difference Equations for Correlation Functions}

In the principal regime the Boltzmann weights of the types
$\check{S}(z)^{i i+1}_{i i+1}$
dominates the others.
Thus in the low temperature limit
$t,q\rightarrow 0$,
only the configuration such that the spin variables take the same value
in the direction from NE to SW
and increase by one in the direction from NW to SE,
is possible.
We call it a configuration of the ground states.
There are $n$ ones labeled by $m\in \bz _n $.
In what follows, we fix one of them (say, $m$)
and define all the correlation functions
in terms of the low-temperature series expansion
(i.e. the formal power series of $t$ and $q$).
Then the lowest order of them comes from
the $m$-th ground state configuration.
Furthermore, any finte order contribution
is derived from
the configurations which differ from
that of the $m$-th ground state
by altering a finite number of spins.
It is equivalent to taking
the GNS representation obtained from the $m$-th ground state
($m$-th GNS representation)
as the Hilbert space.
It is expected that
the correlation function defined in such a way
is an analytic function which has a finite convergence radious
if there exists the phase transition at a finite temperature.

Let us consider the probability

\begin{minipage}{\minitwocolumn}
$$
\hfill
P^{(m)}\left( \begin{array}{l}
z_1 , \cdots , z_N \\
z'_1 , \cdots , z'_N \end{array} \right)^{i_1 , \cdots , i_N }
                                        _{i'_1 , \cdots , i'_N }
:= \sum_{\mbox{config}}
$$
\end{minipage}%
\hspace{\columnsep}%
\begin{minipage}{\minitwocolumn}
\unitlength 1mm
\begin{picture}(50,70)
\put(-7,10){\begin{picture}(101,0)
\put(3,24){$(m)$}
\put(0,10){\vector(1,0){70}}
\put(0,40){\vector(1,0){70}}
\put(10,0){\vector(0,1){50}}
\put(60,0){\vector(0,1){50}}
\put(20,30){\vector(0,1){20}}
\put(19,26){$i_{1}$}
\put(19,51){$z_{1}$}
\put(50,30){\vector(0,1){20}}
\put(49,26){$i_{N}$}
\put(49,51){$z_{N}$}
\put(20,0){\vector(0,1){20}}
\put(19,21){$i'_{1}$}
\put(19,-4){$z'_{1}$}
\put(50,0){\vector(0,1){20}}
\put(49,21){$i'_{N}$}
\put(49,-4){$z'_{N}$}
\put(33,43){$\cdots$}
\put(33,13){$\cdots$}
\end{picture}
}
\end{picture}
\hfill
\end{minipage}
in the inhomogeneous $\bz _n $-Baxter model.
Here $m$ specify the boundary condition
such that the spin on the reference edge
equals to $m$ in the ground state configuration.
The reference edge is the next left
to the one with the spectral parameter $z_1 $.
The symbol of the sum denotes
the statistical sum
over the $m$-th GNS representation.
If we assign the weight $S$ to each vertex,
the statistical sum gives just the probability
in the thermodynamic limit.

The YBE,
the unitarity and the crossing symmetry
permit the following manipulations
%1
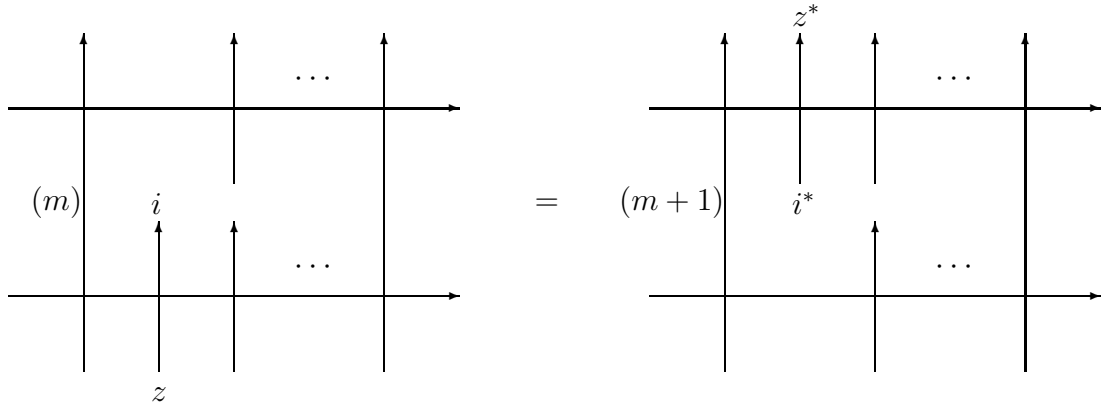
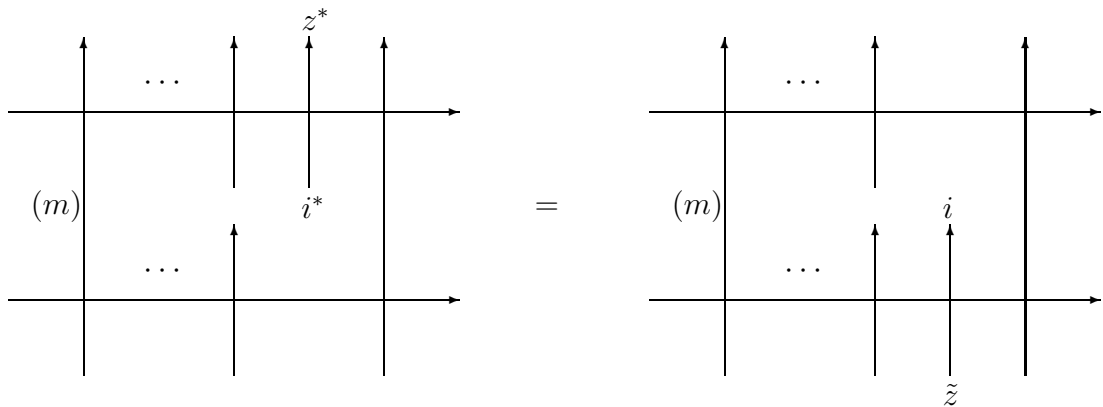
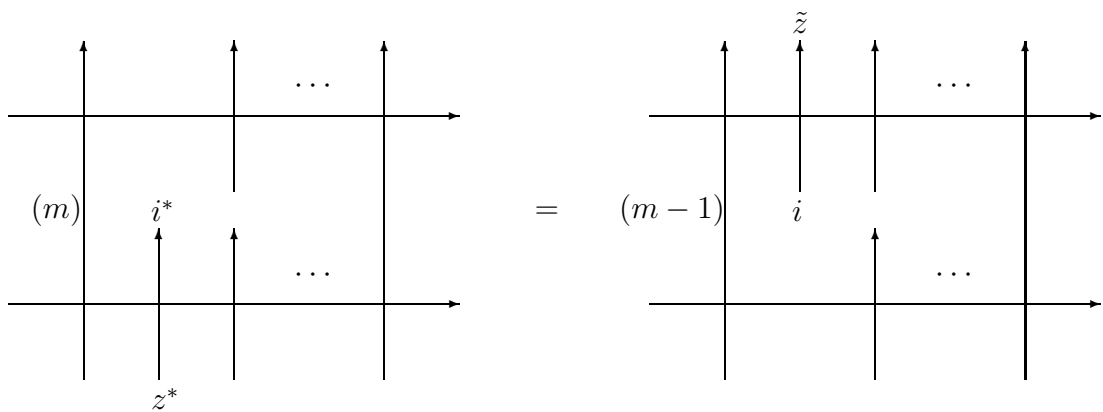
\begin{minipage}{\minitwocolumn}
\unitlength 1mm
\begin{picture}(45,70)
\put(8,10){\begin{picture}(101,0)
\put(3,21.5){$(m)$}
\put(0,10){\vector(1,0){60}}
\put(0,35){\vector(1,0){60}}
\put(10,0){\vector(0,1){45}}
\put(50,0){\vector(0,1){45}}
\put(30,25){\vector(0,1){20}}
\put(40,25){\vector(0,1){20}}
\put(39,21){$i$}
\put(39,46){$z$}
\put(30,0){\vector(0,1){20}}
\put(18,38){$\cdots$}
\put(18,13){$\cdots$}
\put(70,21.5){$=$}
\end{picture}
}
\end{picture}
\end{minipage}%
\hspace{\columnsep}%
\begin{minipage}{\minitwocolumn}
\unitlength 1mm
\begin{picture}(50,70)
\put(8,10){\begin{picture}(101,0)
\put(3,21.5){$(m)$}
\put(0,10){\vector(1,0){60}}
\put(0,35){\vector(1,0){60}}
\put(10,0){\vector(0,1){45}}
\put(50,0){\vector(0,1){45}}
\put(30,25){\vector(0,1){20}}
\put(30,0){\vector(0,1){20}}
\put(40,0){\vector(0,1){20}}
\put(39,21){$i^{*}$}
\put(39,-4){$z^{*}$}
\put(18,38){$\cdots$}
\put(18,13){$\cdots$}
\end{picture}
}
\end{picture}
\end{minipage}
%2
\begin{minipage}{\minitwocolumn}
\unitlength 1mm
\begin{picture}(50,70)
\put(8,10){\begin{picture}(101,0)
\put(3,21.5){$(m)$}
\put(0,10){\vector(1,0){60}}
\put(0,35){\vector(1,0){60}}
\put(10,0){\vector(0,1){45}}
\put(50,0){\vector(0,1){45}}
\put(30,25){\vector(0,1){20}}
\put(20,0){\vector(0,1){20}}
\put(19,21){$i$}
\put(19,-4){$z$}
\put(30,0){\vector(0,1){20}}
\put(38,38){$\cdots$}
\put(38,13){$\cdots$}
\put(70,21.5){$=$}
\end{picture}
}
\end{picture}
\end{minipage}%
\hspace{\columnsep}%
\begin{minipage}{\minitwocolumn}
\unitlength 1mm
\begin{picture}(50,70)
\put(8,10){\begin{picture}(101,0)
\put(-4,21.5){$(m+1)$}
\put(0,10){\vector(1,0){60}}
\put(0,35){\vector(1,0){60}}
\put(10,0){\vector(0,1){45}}
\put(50,0){\vector(0,1){45}}
\put(20,25){\vector(0,1){20}}
\put(19,21){$i^{*}$}
\put(19,46){$z^{*}$}
\put(30,25){\vector(0,1){20}}
\put(30,0){\vector(0,1){20}}
\put(38,38){$\cdots$}
\put(38,13){$\cdots$}
\end{picture}
}
\end{picture}
\end{minipage}
%3
\begin{minipage}{\minitwocolumn}
\unitlength 1mm
\begin{picture}(45,70)
\put(8,10){\begin{picture}(101,0)
\put(3,21.5){$(m)$}
\put(0,10){\vector(1,0){60}}
\put(0,35){\vector(1,0){60}}
\put(10,0){\vector(0,1){45}}
\put(50,0){\vector(0,1){45}}
\put(30,25){\vector(0,1){20}}
\put(40,25){\vector(0,1){20}}
\put(39,21){$i^{*}$}
\put(39,46){$z^{*}$}
\put(30,0){\vector(0,1){20}}
\put(18,38){$\cdots$}
\put(18,13){$\cdots$}
\put(70,21.5){$=$}
\end{picture}
}
\end{picture}
\end{minipage}%
\hspace{\columnsep}%
\begin{minipage}{\minitwocolumn}
\unitlength 1mm
\begin{picture}(50,70)
\put(8,10){\begin{picture}(101,0)
\put(3,21.5){$(m)$}
\put(0,10){\vector(1,0){60}}
\put(0,35){\vector(1,0){60}}
\put(10,0){\vector(0,1){45}}
\put(50,0){\vector(0,1){45}}
\put(30,25){\vector(0,1){20}}
\put(30,0){\vector(0,1){20}}
\put(40,0){\vector(0,1){20}}
\put(39,21){$i$}
\put(39,-4){$\tilde{z}$}
\put(18,38){$\cdots$}
\put(18,13){$\cdots$}
\end{picture}
}
\end{picture}
\end{minipage}
%4
\begin{minipage}{\minitwocolumn}
\unitlength 1mm
\begin{picture}(50,70)
\put(8,10){\begin{picture}(101,0)
\put(3,21.5){$(m)$}
\put(0,10){\vector(1,0){60}}
\put(0,35){\vector(1,0){60}}
\put(10,0){\vector(0,1){45}}
\put(50,0){\vector(0,1){45}}
\put(30,25){\vector(0,1){20}}
\put(20,0){\vector(0,1){20}}
\put(19,21){$i^{*}$}
\put(19,-4){$z^{*}$}
\put(30,0){\vector(0,1){20}}
\put(38,38){$\cdots$}
\put(38,13){$\cdots$}
\put(70,21.5){$=$}
\end{picture}
}
\end{picture}
\end{minipage}%
\hspace{\columnsep}%
\begin{minipage}{\minitwocolumn}
\unitlength 1mm
\begin{picture}(50,70)
\put(8,10){\begin{picture}(101,0)
\put(-4,21.5){$(m-1)$}
\put(0,10){\vector(1,0){60}}
\put(0,35){\vector(1,0){60}}
\put(10,0){\vector(0,1){45}}
\put(50,0){\vector(0,1){45}}
\put(20,25){\vector(0,1){20}}
\put(19,21){$i$}
\put(19,46){$\tilde{z}$}
\put(30,25){\vector(0,1){20}}
\put(30,0){\vector(0,1){20}}
\put(38,38){$\cdots$}
\put(38,13){$\cdots$}
\end{picture}
}
\end{picture}
\end{minipage}
\noindent where $\tilde{z}=z+nw$.

Thus we conclude that the probability
$\displaystyle P^{(m)}\left( \begin{array}{l}
z_1 , \cdots , z_N \\
z'_1 , \cdots , z'_N \end{array} \right)^{i_1 , \cdots , i_N }
                                        _{i'_1 , \cdots , i'_N }$
is given by the following correlation
on the dislocated lattice

\begin{minipage}{\minitwocolumn}
$$
\hspace{-5mm}
F^{(m)}(z'_1 , \cdots , z'_{N},
        z^{*}_N , \cdots , z^{*}_1 )_{
i'_1 , \cdots , i'_{N}, i^{*}_N , \cdots , i^{*}_1 }
= \sum_{\mbox{config}}
$$
\end{minipage}%
\hspace{\columnsep}%
\begin{minipage}{\minitwocolumn}
\unitlength 1mm
\begin{picture}(50,70)
\put(0,11){\begin{picture}(101,0)
\put(3,21.5){$(m)$}
\put(0,10){\vector(1,0){70}}
\put(0,35){\vector(1,0){70}}
\put(10,0){\vector(0,1){45}}
\put(60,0){\vector(0,1){45}}
\put(16,0){\vector(0,1){20}}
\put(15,21){$i'_{1}$}
\put(15,-4){$z'_{1}$}
\put(32,0){\vector(0,1){20}}
\put(31,21){$i'_{N}$}
\put(31,-4){$z'_{N}$}
\put(22,13){$\cdots$}
\put(38,0){\vector(0,1){20}}
\put(37,21){$i^{*}_{N}$}
\put(37,-4){$z^{*}_{N}$}
\put(54,0){\vector(0,1){20}}
\put(53,21){$i^{*}_{1}$}
\put(53,-4){$z^{*}_{1}$}
\put(44,13){$\cdots$}
\end{picture}
}
\end{picture}
\end{minipage}
If we set $z_{\nu }=z'_{\nu } , i_{\nu }=i'_{\nu } (1\leq \nu \leq N)$
in the above expression,
we can get the probability such that
the spins on $N$ successive vertical spins located in the same row
take the values $i_1 , \cdots , i_N $.

We also define the correlations

\begin{minipage}{\minitwocolumn}
$$
\hfill
F^{(m)}(\zeta _1 , \cdots , \zeta _{2N})_{
\iota _1 , \cdots , \iota _{2N}}
= \sum_{\mbox{config}}
$$
\end{minipage}%
\hspace{\columnsep}%
\begin{minipage}{\minitwocolumn}
\unitlength 1mm
\begin{picture}(50,70)
\put(-7,12.5){\begin{picture}(101,0)
\put(3,21.5){$(m)$}
\put(0,10){\vector(1,0){70}}
\put(0,35){\vector(1,0){70}}
\put(10,0){\vector(0,1){45}}
\put(60,0){\vector(0,1){45}}
\put(20,0){\vector(0,1){20}}
\put(19,21){$\iota _{1}$}
\put(19,-4){$\zeta _{1}$}
\put(50,0){\vector(0,1){20}}
\put(49,21){$\iota _{2N}$}
\put(49,-4){$\zeta _{2N}$}
\put(33,13){$\cdots$}
\end{picture}
}
\end{picture}
\end{minipage}
where the set of $\zeta _j $'s is a permutation of
       $z'_1 , \cdots , z'_N , z^{*}_1 , \cdots , z^{*}_N $,
and the set of $\iota _j$'s is that of
       $i'_1 , \cdots , i'_N , i^{*}_1 , \cdots , i^{*}_N $.
Consider the
$V_{\zeta _1 }\otimes \cdots \otimes V_{\zeta _{2N}}$-valued correlators
\begin{equation}
\begin{array}{rcl}
F^{(m)}(\zeta _1 , \cdots , \zeta _{2N})
& =\displaystyle\sum_{\iota _1 , \cdots , \iota _{2N}}
& v_{\iota _1 } \otimes \cdots \otimes v_{\iota _{2N}}
F^{(m)}(\zeta _1 , \cdots , \zeta _{2N})_{\iota _1 , \cdots , \iota _{2N}}
\end{array}
\label{KFm}
\end{equation}
where $V_{z^{*}}=V^{*}_{z} $ and
      $v_{i^{*}}=v^{*}_i $.

The correlators (\ref{KFm})
have the $S$-matrix symmetry:
\begin{equation}
F^{(m)}(\cdots , \zeta _{j+1} , \zeta _j, \cdots )=
\check{S}_{j;j+1}^{V_{\zeta _j }, V_{\zeta _{j+1}}}
F^{(m)}(\cdots , \zeta _j , \zeta _{j+1}, \cdots ).
\label{ax1}
\end{equation}
Furthermore, it satisfies
\begin{equation}
F^{(m\pm 1)}(\zeta _2 , \cdots , \zeta _{2N}, \tilde{\zeta _1 })
=
P_{2N-1;2N} \cdots P_{1;2}
F^{(m)}(\zeta _1 , \zeta _2 , \cdots , \zeta _{2N}).
\label{ax2}
\end{equation}
In the left hand side we should take
$+$ from the double signs
if $\zeta_1 =z'_{\nu }, (1\leq \exists \nu \leq N)$,
$-$
if $\zeta_1 =z^{*}_{\mu }, (1\leq \exists \mu \leq N)$.
The symbol $\tilde{\zeta}$ means $z+nw$ if
$\zeta =z$, $(z+nw)^{*}$ if $\zeta =z^{*}$.
By setting $n=2$
eqs. (\ref{ax1}) and (\ref{ax2})
give those of the eight-vertex model \cite{JMN}.
In \cite{FF}
these are two of axioms that the form factors of
integrable massive models should obey.
As for the derivation of them, see \cite{FF,JMN}.

{}From (\ref{ax1}) and (\ref{ax2}) we obtain the difference equation
\begin{equation}
\begin{array}{rcl}
F^{(m\pm 1)}(\cdots , \tilde{\zeta _j }, \cdots ) & = &
(S^{V_{\tilde{\zeta_j }}, V_{\zeta _{j+1}}}_{j;j+1})^{-1}
\cdots
(S^{V_{\tilde{\zeta_j }}, V_{\zeta _{2N}}}_{j;2N})^{-1}\times \\
{}~ & ~ & \times
S^{V_{\zeta _1 }, V_{\zeta_j }}_{1;j}
\cdots
S^{V_{\zeta _{j-1}}, V_{\zeta_j }}_{j-1;j}
F^{(m)}(\cdots , \zeta _j , \cdots ),
\end{array}
\end{equation}
where the interpretation of the left hand side
is the same as that of (\ref{ax2}).

\section{Spontaneous Polarization}

In this section let us concentrate to
$F^{(m)}(z', z^{*})$.
Since it depends only upon
$u=\exp (-\pi \sqrt{-1} (z-z'))$,
we denote $F^{(m)}(z', z^{*})$ by $F^{(m)}(u)$.
Set
\begin{equation}
G^{(m)}(u):=\sum_{l=0}^{n-1} \omega ^{ml} F^{(l)}(u)_{00^* }.
\end{equation}
Then we have
\begin{equation}
\frac{G^{(m)}(uq^{-n})}
     {G^{(m)}(u)}=
\sum_{j=0}^{n-1} \omega ^{m(j-1)}
(S^{V_{z'}, V_{z}^{*}})^{00^* }_{jj^* }, \label{pol}
\end{equation}
where we use the $\bz _n $-symmetry
$F^{(m)}(u)_{jj^* }=
F^{(m+p)}(u)_{j+p (j+p)^* }$.
Let us solve (\ref{pol}) under the condition
$\lim_{t,q\rightarrow 0} G^{(m)}(1)=\omega ^{-m}$,
which follows from that
$F^{(m)}(z',z^{*})_{jj^{*}}|_{z=z'} \rightarrow \delta^{m+1}_{j}$
in the low temperature limit.

Using the crossing symmetry for $S$-matrix, we get
$$
(S^{V_{z'}, V_{z}^{*}})^{00^* }_{jj^* }=
(S^{V_{z}, V_{z'}})^{j0}_{0j}=
\frac{u^{(n-2)/n}}{\bar{\kappa }(u)}
\frac{(t^{2n}; t^{2n})^{2}_{\infty}}
     {(t^{2}; t^{2})^{2}_{\infty}}
\frac{(q^2 ; t^2 )_{\infty}(t^2 q^{-2} ; t^2 )_{\infty}}
     {(u^2 ; t^{2n})_{\infty}(t^{2n}u^{-2} ; t^{2n})_{\infty}}
\frac{\theta^{(j)}(z-z'+w)}{\theta^{(j)}(z)}.
$$
It follows from the transformation properties for
the theta functions,
\begin{eqnarray*}
\sum_{j=0}^{n-1} \omega ^{m(j-1)}
\frac{\theta ^{(j)}(z-z'+w)}{\theta ^{(j)}(z)} & = &
n\omega ^{-m}
\frac{h((z-m)/n+w)\prod_{l\neq m} h((-z+l)/n)}
     {h(w)        \prod_{l\neq 0} h(l/n)}
\\
{}~ & = &
u^{-(n-2)/n}
\frac{\prod_{l\neq m} (\omega ^l u^{2/n}; t^{2})_{\infty}
                      (t^{2} \omega ^{-l} u^{-2/n} ; t^{2})_{\infty}}
     {\prod_{l\neq 0} (\omega ^l t^2 ; t^2 )_{\infty}
                      (\omega ^{-l} t^2 ; t^2 )_{\infty}} \times \\
{}~ & ~ &
\times
\frac{(\omega ^{-m} q^2 u^{-2/n}; t^{2})_{\infty}
      (t^{2}\omega ^{m} q^{-2}u^{2/n}; t^{2})_{\infty}}
     {(q^2 ; t^2 )_{\infty}(t^2 q^{-2} ; t^2 )_{\infty}},
\end{eqnarray*}
where we use (\ref{fpro}) at the second equality.
Therefore (\ref{pol}) reduces to
\begin{equation}
\frac{G^{m}(uq^{-n})}
     {G^{m}(u)}=
\frac{1}{\bar{\kappa }(u)}
\frac{(\omega ^{-m} q^2 u^{-2/n}; t^{2})_{\infty}
      (t^{2}\omega ^{m} q^{-2}u^{2/n}; t^{2})_{\infty}}
     {(\omega ^{m} u^{2/n}; t^{2})_{\infty}
      (t^{2}\omega ^{-m} u^{-2/n}; t^{2})_{\infty}}.
\end{equation}
Note that the relation
$$
\frac{\varphi (uq^{-n})}{\varphi (u)}=
\frac{1}{\bar{\kappa }(u)},
$$
where
$$
\varphi (u):=g(uq^{n/2})g(u^{-1}q^{n/2}), ~~~~
g(u):=\frac{(q^{3n} u^{-2}; t^2 , q^{2n}, q^{2n})_{\infty}
            (t^2 q^{3n} u^{-2}; t^2 , q^{2n}, q^{2n})_{\infty}}
           {(q^{2+n} u^{2}; t^2 , q^{2n}, q^{2n})_{\infty}
            (t^2 q^{-2+n} u^{2}; t^2 , q^{2n}, q^{2n})_{\infty}}.
$$
Then we obtain
\begin{equation}
G^{(m)}(u)=\omega ^{-m}
           \frac{\varphi (u)}{\varphi (1)}
           \frac{(q^2 ; q^{2})_{\infty}^{2}}
                {(t^2 ; t^{2})_{\infty}^{2}}
\frac{(t^2 \omega ^{m} u^{2/n}; t^{2})_{\infty}
      (t^{2}\omega ^{-m} u^{-2/n}; t^{2})_{\infty}}
     {(q^2 \omega ^{m} u^{2/n}; q^{2})_{\infty}
      (q^{2}\omega ^{-m} u^{-2/n}; q^{2})_{\infty}}.
\end{equation}
The uniqueness of this solution is ensured
under the assumption of the analyticity.

If we define $E^{(m)}(u)=\sum_{j=0}^{n-1} \omega ^j F^{(m)}(u)_{jj^* }$,
then $E^{(m)}(u)=\omega ^m G^{(-1)}(u)$.
Since it is the expectation value of $g$
defined in (\ref{gh}),
the polarization of this model is given as follows:
\begin{equation}
\begin{array}{rcl}
\langle g \rangle ^{(m)} & = & E^{(m)}(u)|_{u=1} \\
{}~ & = &
\omega ^{m+1}
\displaystyle\frac{(q^2 ; q^{2})_{\infty}^{2}}
                  {(t^2 ; t^{2})_{\infty}^{2}}
\displaystyle\frac{(t^2 \omega ; t^{2})_{\infty}
                   (t^{2}\omega ^{-1} ; t^{2})_{\infty}}
                  {(q^2 \omega ; q^{2})_{\infty}
                   (q^{2}\omega ^{-1} ; q^{2})_{\infty}}.
\end{array}
\end{equation}
It reproduces the polarization of the
eight-vertex model conjectured by
Baxter and Kelland \cite{BK} and confirmed in \cite{JMN}
when $n=2$.
Taking the limit
$t\rightarrow 0$, it gives the result of \cite{Ko}
in which the one point function of
the $\frak s \frak l _n $-analog
of the XXZ model is calculated.
We can also obtain
the expectation value of $g^{l}$
\begin{equation}
\langle g^l \rangle ^{(m)}=
\omega ^{l(m+1)}
\displaystyle\frac{(q^2 ; q^{2})_{\infty}^{2}}
                  {(t^2 ; t^{2})_{\infty}^{2}}
\displaystyle\frac{(t^2 \omega ^{l} ; t^{2})_{\infty}
                   (t^{2}\omega ^{-l} ; t^{2})_{\infty}}
                  {(q^2 \omega ^{l} ; q^{2})_{\infty}
                   (q^{2}\omega ^{-l} ; q^{2})_{\infty}}.
\end{equation}

\section{Concluding Remarks}

In this article we study the symmetry of
the correlation function
of the inhomogeneous $\bz _n $-Baxter model
on the dislocated lattice,
which coincide the probability
of the appropriate spin configurations
if one tune the value of
spectral parameters and spin variables.
Such correlations can be characterized
as solutions of the system of difference equations.
By solving the simplest one,
we obtain the spontaneous polarization.
This result includes various ones already obtained
\cite{BK,JMN,Ko}
as special cases.

The next problem is to calculate
the $N$-point function of
the $\bz _n $-Baxter model.
In the trigonometric case
the free field representation
of the $q$-deformed vertex operator
of the quantum affine algebra and
the resultant integral formulae
of $N$-point functions were obtained \cite{JMMN,KQS}.
Since
the elliptic generalization of the
quantum affine algebra has not yet been found,
it is difficult to apply the strategy in \cite{JMMN,KQS}
to the present case.
Thus it is promising to derive and solve
the recursion relation of the correlations.

The critical behavior of this model is
also interesting.
The six-vertex model with $q^{m}=1$,
which is related to minimal models of conformal field theories,
is the critical limit of the eight-vertex model \cite{ESM}.
Refs. \cite{DFJMN,JMMN,IIJMNT} treat
the six-vertex model
and its higher spin analog in the antiferromagnetic regime
($-1<q<0$) because they are based on
the corner transfer matrix method \cite{ESM}.
It is valuable to discuss
the above trigonometric models with $|q|=1$
as the critical limit of
the $\bz _n $-Baxter model.

\section*{Acknowledgement}
I would like to thank
T. Miwa and T. Nakatsu for useful discussion.
This work is partly supported by the Grant-in-Aid for
Scientific Research from the Ministry of Education,
Science and Culture (No. 04-2297).

\end{document}